\documentstyle{article}
\newcommand{\be}{\begin{equation}}
\newcommand{\ee}{\end{equation}}

\begin{document}

\title
{\Large {\bf Assisted Contraction}}
\author{Fabio Finelli \footnote{e-mail: finelli@tesre.bo.cnr.it} \\
I.A.S.F. - Sezione di Bologna, C.N.R. \\ Via Gobetti 101,
40129 Bologna, Italy}
\date{\today}
\maketitle
\begin{abstract}
We consider the dynamics of a contracting universe ruled by two minimally
coupled scalar fields with general exponential potentials. This model
describes string-inspired scenarios in the Einstein frame.
Both background and perturbations can be solved analytically in this
model. Curvature perturbations are generated with a scale invariant
spectrum only for a dust-like collapse, as happens for a single field
model with an exponential potential. We find the conditions for which a 
scale invariant spectrum for isocurvature perturbation is generated.
\end{abstract}

\vskip 0.4cm

\section{Introduction}

String cosmology has been recently arena of new developments. The
Pre-Big-Bang (henceforth PBB) scenario
\cite{Gasperini:1993em,Veneziano:2000pz}, starting
from the low energy string effective action, included a dilaton driven
phase which occurs before the radiation era.
The Ekpyrotic scenario \cite{Khoury:2001wf} uses ideas
from brane cosmology inspired by string theory. According to this
scenario, our universe is a 4-D
orbifold fixed plane in a 5-D Horava-Witten model with a bulk brane moving
towards us in 
the extra space dimension. Both scenarios are alternative ideas to
inflation, and aim to solve the horizon problem with a phase of 
contraction (seen in an auxiliary conformal frame for the PBB or a
physical one in the 4-D effective theory for the Ekpyrotic) rather than a
superluminal expansion.

It seems difficult to obtain a scale invariant
spectrum for curvature perturbations in the simplest single-field
realizations of both these scenarios.  
Curvature perturbations have a white noise spectrum (up to a logarithmic 
factor) in the PBB case during
the dilaton driven phase \cite{BGGMV,DM}, and a vacuum spectrum in the
Ekpyrotic scenario during the brane approach \cite{lyth,BF,hwang,tsuji}.
It has been proposed for this purpose that curvature perturbations could
inherit a scale
invariant spectrum from the growing mode of the Newtonian potential during
the contracting phase \cite{ekpdens} \footnote{According to this idea, the
PBB model would 
have a very infrared spectrum for the Newtonian potential, as 
$k^{-4}$ \cite{ruth}.}.
Unusual matching conditions have been proposed to this purpose
\cite{ekpdens,ruth}, but the debate on the spectrum of
gravitational fluctuations after a bounce is still far from settled
\cite{all}.

In this paper we study isocurvature perturbations arising in multifield
realizations of these string-inspired cosmological models. Indeed, in the
PPB scanario, a scale invariant spectrum was obtained for axion
fluctuations \cite{cew}. We consider a toy model with two
minimally coupled scalar field $\varphi$ and $\chi$ with generic
exponential potential, described by the action:
\be
S = \int d^4 x \sqrt{-g} \left[ \frac{R}{2 \kappa^2} - \frac{1}{2}
\partial_\mu \varphi \partial^\mu \varphi - \frac{1}{2}
\partial_\mu \chi \partial^\mu \chi - V(\varphi,\chi) \right]
\label{action}
\ee
where $\kappa^2 = 8 \pi G \equiv M_{\rm pl}^{-2}$ 
($M_{\rm pl}$ is the reduced Planck mass) and the potential is:
\be
V(\varphi,\chi) = V_1 e^{-\beta \kappa \varphi - \lambda \kappa \chi}
+ V_2 e^{- \alpha \kappa \varphi - \gamma \kappa \chi}
\label{potential}
\ee

Models of this type were studied by Liddle, Mazumdar and Schunck
\cite{assisted} in inflationary context and dubbed {\em assisted
inflation} (see also \cite{malwands,generalized}). 
The idea is that, in the context of exponential potentials 
\cite{lm}, several fields may cooperate to support an inflationary phase
capable to produce density perturbations with an acceptable spectrum,
even if each field would have a potential too steep to make it. 

Exponential potentials similar to (\ref{potential}) are generated in
several particle physics models, as in Kaluza-Klein theories, in 11-D
supergravity models compactified on squashed seven spheres
\cite{hawreall}, or in generalized Scherk-Schwarz compactification of
higher-dimensional Einstein vacuum space-times \cite{green}.

Here instead we study the same action (\ref{action}), but
leading to a Robertson-Walker metric 
\be
ds^2 = g_{\mu \nu} d x^\mu d x^\nu = - dt^2 + a^2(t) d{\bf x}^2 
\label{metric}
\ee
in which the scale factor $a(t)$ slowly contracts:
\be
a(t) \propto (-t)^p \,, \quad t<0 \,, \quad 0 < p < 1
\ee
As we have already said, these contracting space-times are of interest in
string cosmology.
In the Pre Big Bang scenario this contraction
corresponds to a super-inflationary phase in the string frame which
precedes the Big Bang. In the recently proposed Ekpyrotic scenario
\cite{Khoury:2001wf}, this
phase corresponds to a 4-D effective description of the approaching of the
bulk brane to the orbifold we live on.

The 4-D effective action of the Ekpyrotic model includes three fields, two
of which are set to be constant in the original single field model
proposed in \cite{Khoury:2001wf}. 
A simple form for the kinetic part of the 4-D effective action is 
\cite{copeland}:
\be
S_{kin} = \int d^4 x \sqrt{-g} \left[ - \frac{1}{4} 
\partial_\mu \phi \partial^\mu \phi - \frac{3}{4}
\partial_\mu \beta \partial^\mu \beta - 
\frac{c}{2} e^\frac{\beta - \phi}{M_{\rm pl}} 
\partial_\mu z \partial^\mu z \right] \,,
\label{effective}
\ee
where $\phi$ is the dilaton, $\beta$ is the volume modulus and $z$ the
brane modulus \cite{copeland}. Once $\phi$ and $\beta$ are canonically
normalized, and 
$z$ is kept fixed, the kinetic part of the actions (\ref{action}) and
(\ref{effective}) are the same. The kinetic part of the action
(\ref{action}) may also correspond to the case in which there is not a
third brane in the bulk, but the boundary brane is falling on our visible
brane.

According to the PBB scenario, moduli in addition to the dilaton in the
Einstein frame may correspond to radii of the extra dimensions
\cite{cew,buonanno}. The action (\ref{action}) 
would correspond to a modified scenario where both the dilaton and the
moduli have exponential potentials.

The case in which the dilaton or the modulus in the
effective theory of the Ekpyrotic model (\ref{effective}) are
stabilized, leads to a two field background in which the modulus $z$ does 
not have a canonical kinetic term. 
A similar action is obtained in the PBB scenario, by considering the axion
in addition to the dilaton in four dimensions. In the case of vanishing
potential for $z$ or for the axion, 
and exponential potential for the other field, a new exact solution
has been found (see second reference in \cite{BF}). Adiabatic 
and isocurvature perturbations in this
model have been investigated only recently \cite{notari,new}.

The paper is organized as follows. In Section 2 the solution for the
background evolution is presented and in Section 3 the action considered 
is simplified by a rotation in field space. In Section 4 adiabatic and
isocurvature perturbations are studied, following the approach of Gordon
et al. \cite{gordon}. Finally, in Section 5 we conclude.

\section{Two Field Background}

The exact solution occurs for a power-law evolution of the scale factor:
\be
a(\eta) = (- M_{\rm pl} (1-p) \eta)^\frac{p}{1-p} \,,
\label{powerlaw}
\ee
where we are restricting ourselves to the contracting case $ - \infty <
\eta < 0$ and to $0 < p <1$ \footnote{However, all the formula are valid
also
in the case of accellerated expansion $p>1$.}. The following relation
exists among the exponents:
\be
\frac{p}{2} = \frac{(\beta - \alpha)^2 + (\gamma - \lambda)^2}{(\beta
\gamma - \lambda \alpha)^2} \,.
\label{exponents}
\ee

\noindent
The solution for the scalar fields are of logarithmic type 
(we are neglecting integration constants):
\begin{eqnarray}
\varphi (\eta) &=& A \log [ - M_{\rm pl} (1-p) \eta ] \\
\chi (\eta) &=& C \log [ - M_{\rm pl} (1-p) \eta ] \,.
\end{eqnarray}
The coefficients $A \,, C$ are related to the exponents in the following
way:
\begin{eqnarray}
\frac{A}{M_{\rm pl}} &=& \frac{2}{1-p} \, \frac{\gamma - \lambda}{\gamma
\beta - \alpha \lambda} \nonumber \\
\frac{C}{M_{\rm pl}} &=& \frac{2}{1-p} \, \frac{\beta - \alpha}{\gamma
\beta - \alpha \lambda}
\label{relations}
\end{eqnarray}
and 
\be
\frac{V_1 + V_2}{M_{\rm pl}^4} = p \, (3p-1) \,.
\label{potline}
\ee

We now introduce the average field $\sigma$ and its orthogonal field $s$,
by following the formalism of Ref. \cite{gordon} \footnote{See
also the same formalism used as a rotation in
the background field space in the context of assisted inflation
\cite{malwands}. We note that this rotation always leads to a second
field $s$, orthogonal to $\sigma$, which remains constant in time by
definition.}. This method will be particularly useful for the study of
adiabatic and isocurvature perturbations in the next section. The fields
$\sigma$ and $s$ are obtained by a rotation of $\varphi$ and $\sigma$: 
\begin{eqnarray}
{\rm d} \sigma = \cos \theta \, {\rm d} \varphi + \sin \theta \, {\rm
d} \chi \nonumber \\
{\rm d} s = - \sin \theta {\rm d} \varphi + \cos \theta \, {\rm
d} \chi
\label{rotation}
\end{eqnarray}
where 
\begin{eqnarray}
\cos \theta &=& \frac{\dot \varphi}{\sqrt{\dot \varphi^2 +
\dot \chi^2}} = \frac{\dot \varphi}{\dot \sigma} = \frac{A}{M_{\rm pl}}
\frac{1-p}{\sqrt{2p}}
\nonumber \\
\sin \theta &=& \frac{\dot \chi}{\sqrt{\dot \varphi^2 +
\dot \chi^2}} = \frac{\dot \chi}{\dot \sigma} = \frac{C}{M_{\rm pl}}   
\frac{1-p}{\sqrt{2p}} \,.
\label{angle}
\end{eqnarray}
Since both $\varphi$ and $\chi$ scale in the same (logarithmic) way, 
$\theta$ is constant:
\be
\theta = \arctan \frac{C}{A} \,.
\ee
This means that the motion occurs in a straight
line in the field space $(\varphi, \chi)$. 

\section{Rotation in Field Space}

It is interesting to rewrite the action (\ref{action}) in terms of 
the average field $\sigma$ and the orthogonal field $s$ 
\cite{malwands}. This procedure is particularly 
easy for exponential potentials, since it is possible to integrate
directly Eqs. (\ref{rotation}), being $\theta$ constant in time. 
Therefore the action in Eq. (\ref{action}) with the rotated fields is:
\be
S = \int d^4 x \sqrt{-g} \left[ \frac{R}{2 \kappa^2} - \frac{1}{2} \, 
\partial_\mu \sigma \, \partial^\mu \sigma - \frac{1}{2}
\partial_\mu s \, \partial^\mu s - e^{ - \sqrt{\frac{2}{p}} \kappa \sigma}   
\bar{V} (s) \right]
\label{action2}
\ee
with
\begin{eqnarray}
\bar{V} (s) &=& V_1 e^{-\kappa s (\lambda \cos \theta - \beta \sin
\theta)} + V_2 e^{-\kappa s (\gamma \cos \theta - \alpha \sin \theta)}
\nonumber \\
&=& M_{\rm pl}^4 p (3p-1) \left[ x e^{\sqrt{p/2} d \kappa s (1-x)} + 
(1-x) e^{- \sqrt{p/2} d \kappa s x} \right]
\,,
\label{vesse}
\end{eqnarray}
where we have neglected the integration constants in integrating 
Eq. (\ref{rotation}) (which would just modify $V_1, V_2$), we have
defined $d = \gamma \beta - \alpha \lambda$ and introduced 
\be
x = \frac{V_1}{p (3p - 1) M_{\rm pl}^4} \,.
\label{xy}
\ee
By definition 
$s$ is constant, which means that $s$ sits in the extremum of the
potential 
(\ref{vesse}), $s_{\rm min} = 0$. In Fig. 1 we show the potential 
$\bar{V} (s)$ for $x=0.1$ and its dependence on $p$ and $d$.
It is instructive to study the potential $\bar{V} (s)$ close to the
extremum, by Taylor expanding: 
\be
\bar{V} (s) = M_{\rm pl}^4 p (3p-1) \left\{ 1 + \frac{s^2}{2 
M_{\rm pl}^2} x (1-x) 
\left[ (\beta - \alpha)^2 + (\gamma - \lambda)^2 \right] + {\cal O} (s^3)
\right\} \,.
\label{barpotexp}
\ee

\begin{figure}
\vspace{5.5cm}
\includegraphics{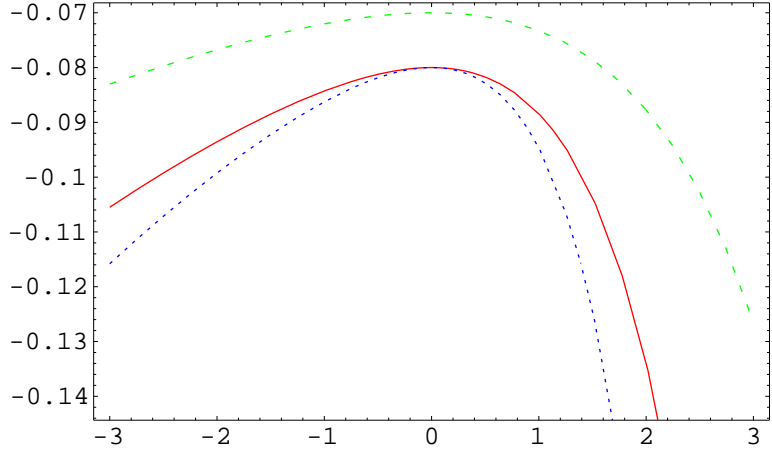}
\caption{The potential $\bar{V} (s)$ (in units of $M_{\rm pl}^4$) as a
function of $s/M_{\rm pl}$ for
$p=0.2$, $d=4$ (solid line) and $d=5$ (dotted line), and for $p=0.1$,
$d=4$ (dashed line). For all the curves $x=0.1$.}
\label{pippo}
\end{figure}

For a better understanding of the form of the potential in (\ref{action2}),  
it is also possible to write the problem in a string frame conformally 
related to the metric $g_{\mu \nu}$ used in (\ref{metric}). 
By following Ref. \cite{malwands}, we get (by neglecting boundary terms):
\be
\tilde{S} = \int d^4 x \sqrt{-\tilde{g}} e^{-\tilde{\sigma}} 
\left[  \frac{1}{2 \kappa^2} \left( \tilde{R} - \omega
\tilde g^{\mu \nu} 
\partial_\mu \tilde{\sigma} \partial_\nu \tilde{\sigma} \right) -
\frac{1}{2} \tilde g^{\mu \nu}
\partial_\mu s \partial_\nu s - \bar{V} (s) \right] \,,
\label{action3}
\ee
where $\tilde{\sigma} = - \sqrt{2/p} \kappa \sigma$ and
\be
\tilde{g}_{\mu \nu} = e^{\tilde{\sigma}} g_{\mu \nu} \, \quad \, 
\omega = \frac{p - 3}{2} \,.
\ee
We note that for $p=1$ the kinetic term of $\tilde \sigma$ is just the
usual one considered in the low-effective action for string theory in four
dimension.

\section{Perturbations}

We now consider the perturbations around the background defined 
in the previous section. We shall focus directly on 
adiabatic and isocurvature fluctuations. We use the curvature
perturbation $\zeta$ \cite{Lyth:1985gv,Mukhanov:1992me} to follow the
evolution of the adiabatic mode. In the longitudinal gauge for metric
perturbations:
\be
ds^2 = - (1 + 2 \Phi (t, {\bf x})) dt^2 + a(t)^2 (1 - 2 \Phi (t, {\bf x}))
d{\bf x}^2 
\ee
the variable $\zeta$ is \cite{Mukhanov:1992me}:
\be
\zeta = - \frac{H}{\dot H} (\dot \Phi + H \Phi) + \Phi = 
\frac{H}{\dot \sigma} \delta \sigma + \Phi \,,
\ee
where in last equality $\zeta$ is expressed as function of the
fluctuation of the average field \cite{gordon}. The time evolution of
$\zeta$ is \cite{gb,Finelli:2000ya}:
\be 
\dot \zeta = - \frac{H}{\dot H} \frac{\nabla^2}{a^2} \Phi 
+ \frac{H}{2} \left( \frac{\delta \varphi}{\dot \varphi} -
\frac{\delta \chi}{\dot \chi} \right) \frac{d}{dt} \left(
\frac{\dot \varphi^2 - \dot \chi^2}
{\dot \varphi^2 + \dot \chi^2} \right)
= - \frac{H}{\dot H} \frac{\nabla^2}{a^2} \Phi + \frac{2 H}{\dot \sigma}
\dot \theta \delta s
\ee

As follows from Ref. \cite{gordon} curvature and isocurvature 
perturbations are decoupled when $\dot \theta = 0$. Therefore, in the case 
of exponential potentials curvature $\zeta$ and isocurvature $\delta s$ 
perturbations are a particular rotation in the space of perturbations 
which diagonalize the system of equations of motion. 
The equation for the Fourier mode of $\zeta$ is:
\be
\ddot \zeta_k + (3 H - 2 \frac{\dot H}{H} + \frac{\ddot H}{\dot H}) 
\dot \zeta_k + \frac{k^2}{a^2} \zeta_k = 0
\label{zetaeq}
\ee
For power-law evolution of the scale factor as in Eq. (\ref{powerlaw}),
the damping term is simply $3 H \dot \zeta$ and therefore Eq.
(\ref{zetaeq}) 
is the equation for a massless minimally coupled scalar field. 
In the single field case $\zeta$
satisfies the same equation \cite{lyth}. 
For an evolution of the scale factor 
given by Eq. (\ref{powerlaw}), the solution for $\zeta$ is 
\cite{BF}:
\be
\zeta_k = \frac{1}{2 a M_{\rm pl}} \sqrt{-\frac{\pi p \eta}{2}}
H_{|\nu_\zeta|}^{(2)}
(-k\eta)
\,,
\label{zetasol}
\ee
where $H^{(2)}_{|\nu|}$ is the Hankel function of the second kind and 
the spectral index $\nu$ is given:
\be
\nu_\zeta = \frac{1}{2} \frac{1-3p}{1-p} \,.
\label{adindex}
\ee
A scale invariant spectrum results only for a dust-like collapse, 
i.e. $p=2/3$ \cite{BF}. For an extremely slow contraction, 
i.e. $p \sim 0$, as suggested in the Ekpyrotic scenario \cite{Khoury:2001wf}, 
$\zeta$ has a blue spectrum \cite{lyth,BF,hwang,tsuji}.

We now study the solution for the isocurvature perturbations. The equation
for $\delta s_k$ with $\dot \theta=0$ can be rewritten as \cite{gordon}:  
\be
(a \delta s_k)'' + \left[ k^2 + a^2 V_{ss} - \frac{a''}{a} \right] 
(a \delta s_k) = 0 \,.
\ee
where 
\be
V_{ss} = \sin^2 \theta \, V_{\varphi \varphi} - \sin 2 \theta \, 
V_{\varphi \chi} + \cos^2 \theta \, V_{\chi \chi}
\label{vsecond}
\ee
\be
\frac{a''}{a} = p \frac{2p-1}{(1-p)^2 \eta^2} \,.
\ee
The solution for $\delta s_k$ is:
\be
a \delta s_k = \left( - \frac{\pi \eta}{4} \right)^{1/2} H_{|\nu_s|} (-k
\eta) 
\ee
where $\nu_s$ can be obtained from:
\be
\frac{\nu_s^2 - 1/4}{\eta^2} = \frac{a''}{a} - a^2 V_{ss} \,.
\label{spectralindex}
\ee 
By inserting Eqs. (\ref{powerlaw}-\ref{angle}) in Eq. (\ref{vsecond}) one
gets:
\be
a^2 V_{ss} = \frac{V_1 V_2}{M_{\rm pl}^8 (3p-1)} 
\frac{d^2}{2 (1-p)^2 \eta^2}
\label{masseff}
\ee
By using Eq. (\ref{potline}), it is useful to rewrite 
this last equation as:
\begin{eqnarray}
a^2 V_{ss} &=& p^2 d^2 x(1-x) 
\frac{3p-1}{2 (1-p)^2 \eta^2} \nonumber \\
&=& p \left[ (\beta - \alpha)^2 + (\gamma - \lambda)^2 \right] x(1-x)
\frac{3p-1}{(1-p)^2 \eta^2}
\label{masseff2}
\end{eqnarray}

The effective mass for the isocurvature perturbations is therefore zero
when $V_1$ or $V_2$ vanish. This means that for a potential which is just
a product of exponentials of the single fields, adiabatic and
isocurvature fluctuations are decoupled, but generated with the same
spectrum, produced by the curvature term $a''/a$. The same thing
occurs for a free theory ($V_1=V_2=0\,$, $p=1/3$), a result which is
already known in the PBB scenario \cite{cew}. We also note
that the effective mass in Eq. (\ref{masseff2}) agrees with the curvature
of the potential in Eq. (\ref{barpotexp}). 
For $0 < x < 1$ the
potential $V$ around $s=0$ is concave (convex) for 
$0 < p <1/3$ ( $1/3 < p < 1$ ). 

We now consider the case for which the spectrum of isocurvature
perturbations is scale invariant, i.e. when the following condition is
satisfied:
\be
a^2 V_{ss} - \frac{a''}{a} = - \frac{2}{\eta^2} \,.
\ee
By solving the second order equation for $x$ deriving from the above
equation we find:
\be
x_{1,2} = \frac{1}{2} \pm \sqrt{\frac{1}{4} + \frac{4}{p^2 d^2 (3p - 1)}
(1 - \frac{3}{2}p)} \,.
\label{xsolution}
\ee

In what follows we shall discuss some particular cases.
 
\vspace{1cm}
\noindent
{\bf CASE A: $|\gamma| \,, |\beta| \gg |\lambda|, |\alpha|$.}

\vspace{.5cm}

It is useful to rewrite Eqs. (\ref{relations}) with the help of Eqs.
(\ref{angle}):
\begin{eqnarray}
\gamma &=& \lambda + \sqrt{p \over 2} (\gamma \beta - \alpha \lambda)
\cos \theta
\nonumber \\
\beta &=& \alpha + \sqrt{p \over 2} (\gamma \beta - \alpha \lambda)
\sin \theta
\end{eqnarray}
We now work in the regime in which $|\gamma| \,, |\beta| \gg
|\lambda|, |\alpha|$, which leads to:
\begin{eqnarray}
\beta &\sim& \sqrt{2 \over p} \frac{1}{\cos \theta}
\nonumber \\
\gamma &\sim& \sqrt{2 \over p} \frac{1}{\sin \theta} \,.
\end{eqnarray} 
If also $p \sim 0$ we have $|\gamma| \,, |\beta| \gg 1$. 
Under these two conditions it is easy to show that 
$a^2 V_{ss} \sim - 2 / \eta^2$ and therefore 
\be
\nu_s^2 = \frac{9}{4} - \sqrt{\frac{p}{2}} \left( \frac{\alpha}{\cos
\theta} +
\frac{\lambda}{\sin \theta} \right) + {\cal O} (p) \,, 
\label{casouno}
\ee
i.e. the spectrum of isocurvature perturbations is nearly scale
invariant, with a tilt of the order of $\sqrt{p}$. 

By symmetry, an analog result holds in the opposite case, i. e. $|\gamma|
\,, |\beta| \ll |\lambda|, |\alpha|$, when $p \sim 0$. In this case the
spectral index is the same of Eq. (\ref{casouno}), with $\alpha$ and
$\lambda$ replaced with $\beta$ and $\gamma$, respectively.

\vspace{1cm}
\noindent
{\bf CASE B: $\lambda = \alpha = 0$}.

\vspace{.5cm}

In this case we get the following result for the spectral index in 
Eq. (\ref{spectralindex}):
\be
\nu_s = \frac{3}{2} \frac{\sqrt{(1/3 - p)(3-p)}}{1-p} \,.
\label{sindex2}
\ee
The same result was found in \cite{malwands}. 
We note that the spectral index does not depend on $\theta$, i.e. on how
the total energy is shared between the two fields.
For $1/3 < p <1$, $\nu_s$ is imaginary, which means that isocurvature
perturbations are exponentially dumped. For $0 < p < 1/3$, $\nu_s$ is
real, and in the case of a very slow contraction $p \sim 0$ we get a
nearly scale-invariant spectrum, slightly blue-tilted:
\be
\nu_s = \frac{3}{2} - p + {\cal O} (p^2) \,.
\label{caseb}
\ee

\vspace{1cm}
\noindent
{\bf CASE C}: $p = \frac{1}{2}$.

\vspace{.5cm}

\begin{figure}
\vspace{5.5cm}
\includegraphics{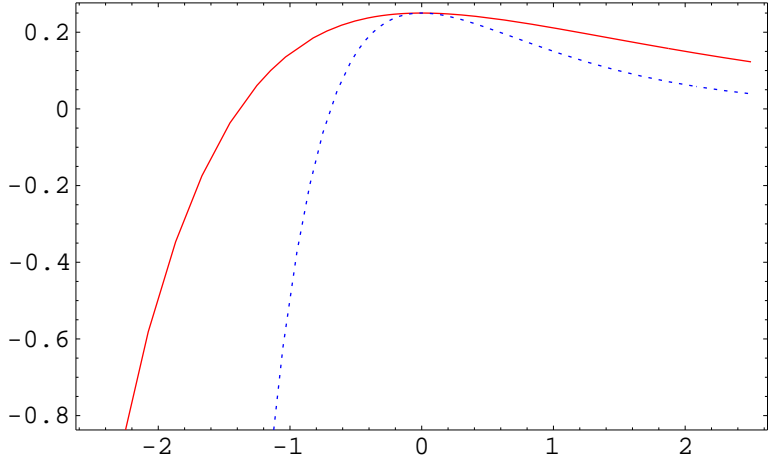}
\caption{The potential $\bar{V} (s)$ (in units of $M_{\rm pl}^4$) as a
function of $s/M_{\rm pl}$ for $p=1/2$ case, with 
$d=1$ (solid line) and $d=2$ (dotted line). For both curves $x=2$.}
\label{pluto}
\end{figure}

In this case we have a radiation type contraction, in which the curvature 
term $a''/a$ is zero. 
For $p = 1/2$ a scale invariant spectrum
for isocurvature perturbations emerges only when $V_1$ and $V_2$ have
opposite signs ($x < 0$ or $x > 1$). We show in Fig. (\ref{pluto}) the
shape of the potential $\bar{V} (s)$ for $x=2$ for two possible values of
$d$.

\begin{figure}
\vspace{6cm}
\includegraphics{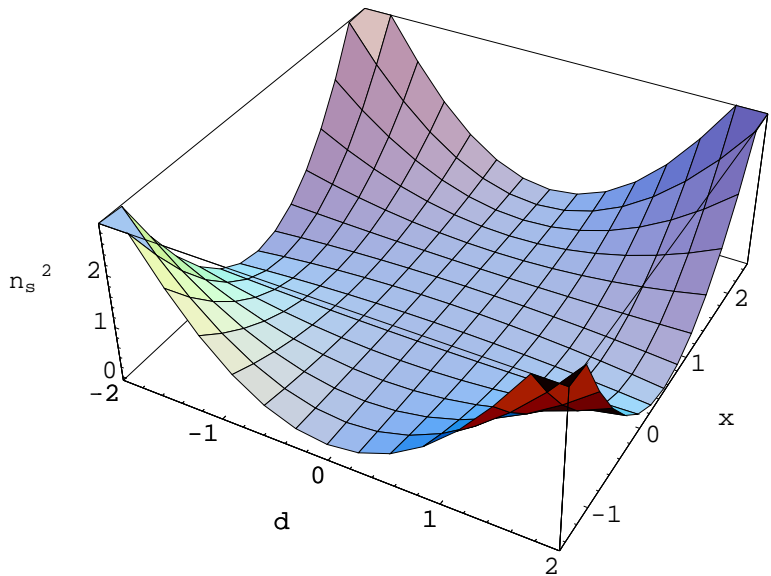} 
\caption{The spectral index for isocurvature perturbations in the case
$p=1/2$ as a function of $d$ and $x$.}
\label{fig1}
\end{figure}

The 3-D plot of $\nu_s^2$ as a function of $x$ and $d$
is presented in Fig. (\ref{fig1}).
As follows from the general case treated in Eq. (\ref{xsolution}), 
$\nu_s = 3/2$ emerges when
\be
x_{1,2}^{p=1/2} = \frac{1}{2} \pm \sqrt{\frac{1}{4} + \frac{8}{d^2}}
\ee

\section{Discussions and Conclusions} 

We have discussed a scenario with a contracting universe driven by 
two minimally coupled scalar fields with a generic exponential 
potential. This model allows exact solutions for background and 
cosmological perturbations and is relevant in string cosmology 
\cite{Gasperini:1993em,Veneziano:2000pz,Khoury:2001wf}, studied 
in the Einstein frame. The corresponding 
solution with an expanding universe has been considered in the framework of 
assisted inflation \cite{assisted}.

The action considered in Eq. (\ref{action}) has a more appealing form by 
rewriting it as a function of the average and the orthogonal field, as in 
Eq. (\ref{action2},\ref{action3}). 
A similar action was considered in the expanding case 
in {\em soft inflation} models \cite{soft}. 

As already mentioned, for the action (\ref{action},\ref{potential})
exact solutions are possible for 
adiabatic and isocurvature perturbations, since these are decoupled. 
In the inflationary context, adiabatic and isocurvature perturbations are 
nearly scale invariant for the same condition, i.e. $p >> 1$ 
\cite{malwands}, as follows from Eqs. (\ref{adindex},\ref{sindex2}). 
The reason being that both adiabatic and isocurvature perturbations 
are amplified by the curvature term $a''/a$ in the range $p >> 1$ 
\footnote{The effective mass for isocurvature perturbations is suppressed by 
$p$ with respect to the curvature term $a''/a$ for $p>>1$, as it can be seen 
from Eqs. (\ref{xy},\ref{masseff2}).}. Instead, during a contraction, 
isocurvature perturbation can develop a 
scale invariant spectrum for $p \sim 0$, while curvature perturbation 
cannot. The curvature term $a''/a$ is suppressed by $p$ and  
ineffective for the amplification of $\zeta$, while the effective mass for 
isocurvature perturbation is negative, and the generation of a scale 
invariant spectrum for the latter is possible.

We have shown that the mechanism to produce a scale invariant
spectrum based on tachyonic instability, rather than the geometric effect 
due to the curvature term $a''/a$, proposed in \cite{KKL}, 
could really work in a multifield scenario. 
This result is also attractive from
the particle physics point of view, since tachyonic instability occurs in
symmetry breaking phenomena, mostly implemented with two or more fields.
In the single field case, the negative effective mass, originating from
the negative exponential potential, is balanced by the metric contribution
(indeed, gravity is ruled by this single field) and 
a scale invariant spectrum for curvature perturbation cannot be generated 
\cite{lyth,BF}, at least for exponential potentials.
In the multifield case, this result is confirmed by the equation of motion
for the adiabatic perturbation (\ref{zetaeq}), which is the same of a
single field one for power-law expansion.
For the isocurvature perturbation $\delta s$ this relation between its
effective potential and the metric contribution is lost. A negative
effective mass could really survive and lead to an infrared spectrum,
as shown here. 

We think this isocurvature mechanism works in a generic way, 
and it does not occur only with the exponential potential
we started from. For the potential (\ref{potential}), $s$ is exactly frozen, 
but one can guess that the result will not be very different when $s$ 
slowly moves, as it happens in the slow-rollover regime for the inflaton.
Indeed, from Eq. (\ref{barpotexp}) the potential around the extremum 
is very simple both in the case B:
\be
\bar{V} (s) \simeq - p \, M_{\rm pl}^4 - M^2_{\rm pl} s^2 \,.
\label{one}
\ee
and in the case C:
\be
\bar{V} (s) \simeq \frac{M_{\rm pl}^4}{4} \left( 1 - \frac{s^2}{M_{\rm
pl}^2} \right) \,.
\label{two}
\ee 

We have concentrate our attention on contractions for which the curvature 
$a''/a$ is minimal, i.e. $p \sim 0$ and $p \sim 1/2$. 
In our opinion, these types of contractions would alleviate (but not
solve) the problem of
amplifying classical inhomogeneities simultaneously with quantum
fluctuations \cite{lectureslinde}. Initial curvature perturbations are
almost inert because the curvature is almost zero, while isocurvature
perturbations are amplified by tachyonic instability. 

It remains to see how and when the scale invariant spectrum of isocurvature 
perturbations is transferred to the adiabatic component. For exponential
potential adiabatic and isocurvature perturbations are decoupled, but for 
potentials like in Eqs. (\ref{one},\ref{two}) are coupled (weakly if
$s$ slowly rolls). Of course, a transfer could also occur during the
bounce - or the graceful exit - or during a stabilization mechanism, 
but this possibility remains to be investigated in detail. 

\vspace{1cm}

{\bf Acknowledgements}

\noindent
It is a pleasure to thank Robert Brandenberger for several
important suggestions and useful comments on the manuscript, and Ruth
Durrer for interesting discussions.

\end{document}